\documentclass[12pt,a4paper]{article}
\usepackage{psfig}
\renewcommand\thesection{\arabic{section}.}
\renewcommand\thesubsection{\thesection\arabic{subsection}.}
\renewcommand\thesubsubsection{\thesubsection\arabic{subsubsection}.}
\renewcommand\section[1]{\vspace{\topsep}\vspace{\partopsep}
\renewcommand{\topfraction}{0.99}
\renewcommand{\bottomfraction}{0.99}
\renewcommand{\textfraction}{0.01}
\renewcommand{\floatpagefraction}{0.95}

\refstepcounter{section}
{\par  \noindent\normalsize\bfseries \thesection
\hspace{1em}#1\vspace{\topsep}\par\noindent}}

\newenvironment{refs}
{\vspace{\topsep}\vspace{\partopsep}
{\par \noindent\normalsize\bfseries  References
\vspace{-\topsep}\par\noindent}
\setlength{\parindent}{-5mm}
\begin{list}{}{\topsep 0pt \partopsep 0pt \itemsep 0pt \leftmargin 5mm
\parsep 0pt \itemindent -5mm}}
{\end{list}}

\renewcommand\subsection[1]{
\refstepcounter{subsection}
{\par \protect\vspace{\topsep}\vspace{\partopsep}
 \noindent\normalsize\bfseries \slshape \thesubsection
\hspace{1em}#1\par \noindent}}

\renewcommand\subsubsection[1]{
\refstepcounter{subsubsection}
{\par \protect \vspace{\topsep}\vspace{\partopsep}
\noindent\normalsize \slshape \thesubsubsection
\hspace{1em}#1\par \noindent}}

\newfont{\sansb}{cmssbx10}
\newfont{\sans}{cmss10}
\begin{document}
\noindent\hfill{EFI 97-52}
\begin{center}
{\large \bf The STACEE Project\\}

\vspace{5mm}

{Ren{\' e} A. Ong and Corbin E. Covault\\}
{\sl 
Enrico Fermi Institute, University of Chicago, Chicago, IL 60637, USA\\}

\vspace{1mm}

{(For the STACEE Collaboration)}

\vspace{5mm}

{To appear in Proc. Towards a Major Atmospheric Cherenkov Detector-V
(Kruger Park)}

\end{center}

\begin{abstract}
The Solar Tower Atmospheric Cherenkov Effect Experiment (STACEE)
is designed to explore the gamma-ray sky
between 20 and 250 GeV using the atmospheric Cherenkov technique.
STACEE will use large solar heliostat mirrors to
reflect Cherenkov light created in gamma-ray air showers to
secondary mirrors on a central tower.
The secondary mirrors image this light onto photomultiplier tube
cameras that are read out by fast electronics.
Here we outline the important features of the STACEE design.
We present an overview of the experimental site, describe the
method of heliostat selection and control, and discuss the
current designs for the detector components, including
the secondary mirror structures, cameras, and electronics.
\end{abstract}
\setlength{\parindent}{1cm}
\sloppy

\section{Introduction}

There is a window in the gamma-ray spectrum which has
yet to be systematically explored by any telescope.
Current state-of-the-art atmospheric Cherenkov telescopes have
energy thresholds of 250 GeV or greater.
Conversely, the EGRET experiment on the Compton Gamma Ray Observatory
detects few astrophysical photons above 20 GeV.
Exploring the gamma-ray window between 20 and 250 GeV is a primary
goal of STACEE.
A more complete discussion of the scientific motivation
for exploring this window can be found elsewhere
(Ong 1997).

To first order, the energy threshold of an atmospheric Cherenkov
telescope is limited by night sky fluctuations and
scales inversely with the square root of the telescope mirror collection
area.
The most straightforward way to achieve
a low gamma-ray 
energy threshold is to use a very large mirror area.
Two approaches to significantly increasing mirror area have been
considered.
The first is to build large (e.g. 10 m diameter) mirrors from scratch.
In this way the mirrors can be custom-built for Cherenkov astronomy
but the costs will be relatively high.
A second approach is to make use of large mirror facilities that
already exist.
More than a decade ago, it was pointed out that large solar mirrors
(heliostats) could be used as the primary collector in
an atmospheric Cherenkov telescope (Danaher {\em et al.} 1982).
Later, a design incorporating a secondary optic was suggested
by members of our group (T{\" u}mer {\em et al.} 1990).
In the past three years, the CELESTE (Par{\' e} 1997 and
Qu{\' e}bert 1997) and STACEE 
(Chantell {\em et al.} 1997, 
Covault {\em et al.} 1997) groups have carried out tests
at solar mirror facilities to demonstrate that low energy 
Cherenkov telescopes can in fact be constructed at these locations.
Both experiments have been funded and are now under construction.
A third experiment called GRAAL (Plaga 1997) is under development.
The designs of CELESTE and STACEE are similar in many respects,
but here we concentrate on the issues relevant for STACEE.

The STACEE collaboration explored the possibility of 
using the heliostat mirrors of the Solar Two power plant
near Barstow CA, USA.  
Initial prototype work at the Solar Two site led to the first
detection of atmospheric Cherenkov radiation at such a facility
(Ong {\em et al.} 1996).
Solar Two is currently an operating power plant, and a more favorable
U.S. location is the research heliostat field of the National
Solar Thermal Test Facility (NSTTF) at Sandia National Laboratories
in Albuquerque NM.

\begin{figure}
\centerline{\psfig{file=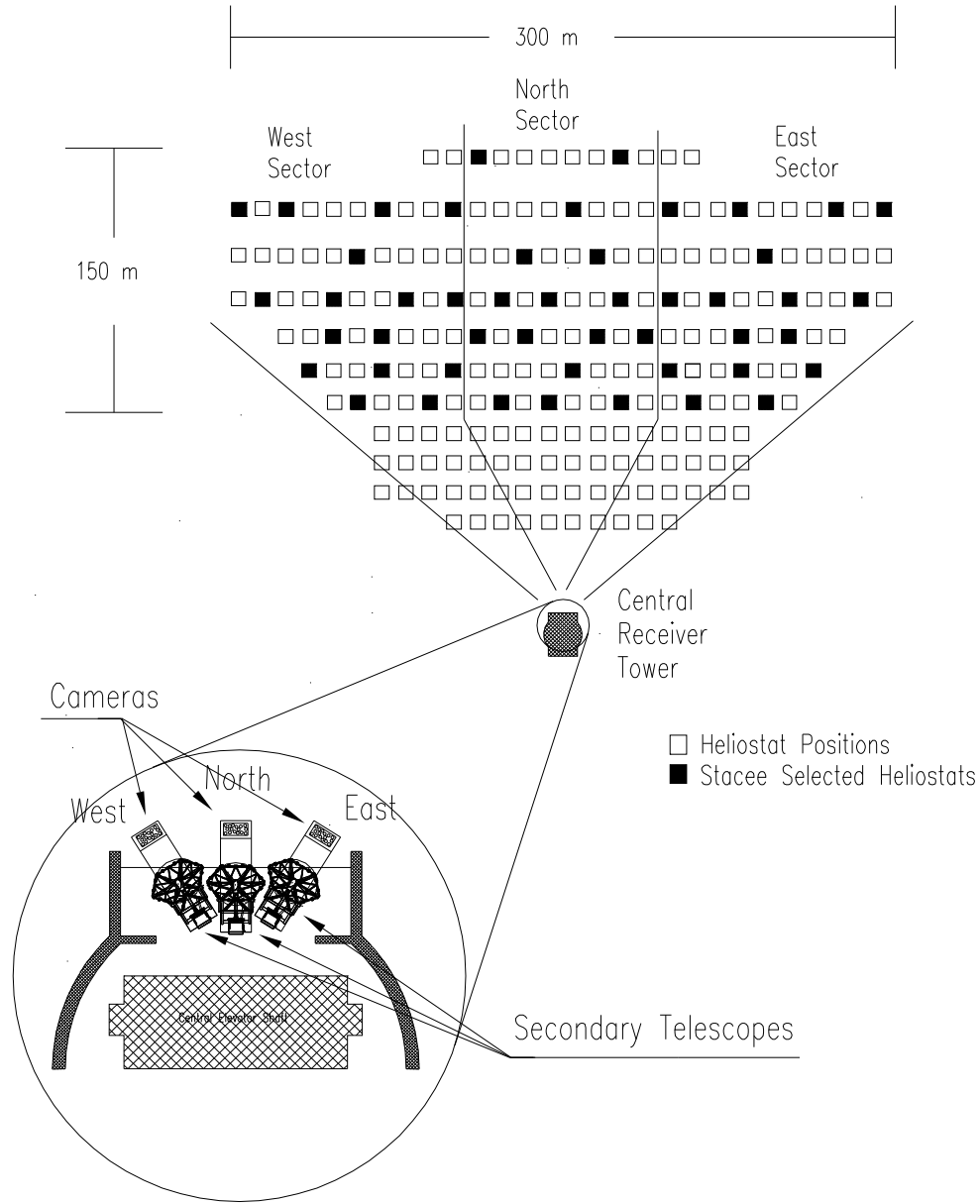,width=14.0cm}}
\caption{Plan view of the STACEE telescope.
STACEE uses 48 heliostat mirrors of the National Solar Thermal
Test Facility near Albuquerque NM, USA (upper drawing).
Each heliostat has an area of 37$\,$m$^2$ and the distance
between heliostat centers is $\sim 12.5\,$m.
The heliostats reflect Cherenkov radiation onto three secondary mirrors
located on a central tower (lower drawing).
The secondary mirrors image the light onto photomultiplier tube cameras.}
\end{figure}

The overall design of STACEE is shown in Figure~1.
Cherenkov light created in gamma-ray air showers 
reaches the ground with a broad lateral extent 
$\sim 250\,$m across.
The NSTTF consists of 220 heliostat mirrors 
(each 37.2$\,$m$^2$ in area).
STACEE uses 48 heliostats for a total reflector surface of
$1785\,$m$^2$.
The heliostats reflect the Cherenkov light onto secondary mirrors on
a central tower.
Three $2\,$m diameter secondary mirrors 
viewing different parts of the heliostat field are used.
Light reflected from each secondary mirror is imaged onto
a photomultiplier tube (PMT) camera.
Each heliostat is viewed by a single PMT.

In this paper, we describe the essential features of the STACEE
design.
We first discuss the NSTTF site itself.
We then outline
the essential hardware components of STACEE, including the:
\begin{enumerate}
\item heliostats,
\item secondary mirror structures,
\item cameras, and
\item electronics.
\end{enumerate}

\section{The Site}

The NSTTF is located approximately 15 km southeast of Albuquerque 
at an altitude of 1700 m above sea level.
The
site is dry and has excellent sky clarity.
Weather patterns accumulated over the last thirty years
predict an average precipitation of 13.4 cm per year
and a yearly average of 170 fully clear days and 111 mostly clear days
(NOAA 1997).
We have measured the intensity of night sky background light
at the site and have found an average value of
$(4.3 \pm 0.9) \times 10^{12}$ photons m$^{-2}$ sec$^{-1}$ sr$^{-1}$
over the wavelength range spanned by a typical bi-alkali
PMT (300-600 nm).

The NSTTF is located at Sandia National Laboratories on Kirtland 
Air Force Base.
The facility is fully supported and operated by the U.S. Department
of Energy (DOE).
As such, the DOE maintains the heliostat field, heliostat control tower,
and solar tower.
Heliostat repair is done as a matter of course.
Since the NSTTF is a research facility and 
not an operating power plant,
there is little potential for interference between STACEE work and
solar energy research.
We are able to schedule our observations well in advance with
minimal interruption from other activities at the site.

%fuj
\newpage

\section{Heliostats}

Heliostat mirrors at the NSTTF consist of 25 square facets 
made of rear-surfaced glass.
Facets for selected
heliostats are aligned and focused using a television camera
mounted at the top of the central tower.   Each heliostat mirror 
can be positioned
in azimuth and elevation with 13-bit precision
(0.05$^\circ$). The optical and pointing performance of the
heliostats is found to be satisfactory for atmospheric Cherenkov astronomy
and has been
described elsewhere (Covault 
{\em et al.} 1997, Chantell
{\em et al.} 1997).

\subsection{Heliostat Selection}

Of the 220 heliostats at the NSTTF site, we have selected 48
heliostats to be used for STACEE (16 heliostats for each of three
secondary mirrors).  Ideally, heliostats are selected so as to provide
uniform coverage over the entire area of the field.  However,
selected heliostats must also not crowd each other in the focal
plane of the secondary mirror.  We also select heliostats to minimize
off-axis aberrations.  Finally, field-of-view requirements prevent
consideration of heliostats that are too close to the central tower.

\begin{figure}
\centerline{\psfig{file=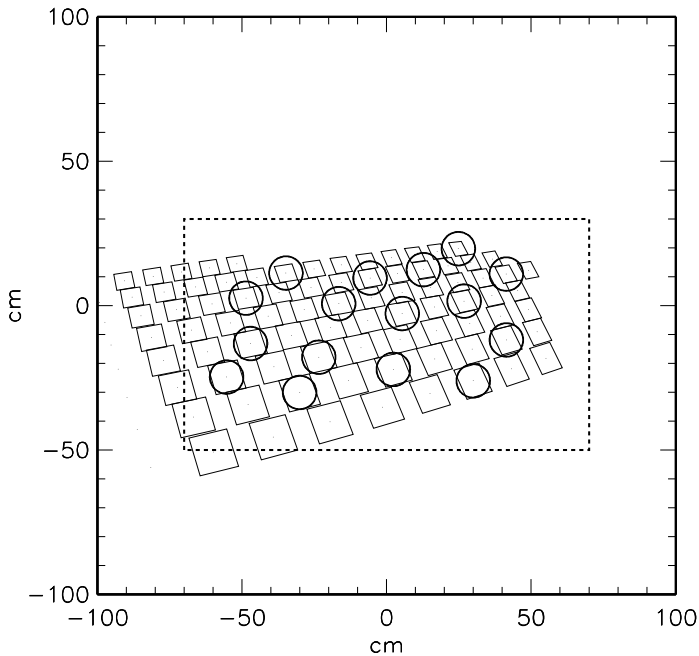,width=10.0cm}}
\caption{View of northeastern corner of NSTTF heliostat field
as seen from the position of the secondary mirror.
Camera outline (rectangle) and position of PMT entrance apertures
(circles) at focal plane are indicated.}
\end{figure}

Figure~2 shows the view from the tower into the northeastern corner
of the heliostat array corresponding to one of the three secondary
mirrors.  This view is imaged onto the focal plane of the mirror.  The
rectangular border indicates the boundary of the camera. Thick
circles indicate the position of each PMT can entrance aperture for
the 16 selected heliostats in this part of the array.  Note that
in actual operation, non-selected heliostats will be turned edge-on
and thus will not reflect light up to the
secondary mirror.

\subsection{Heliostat Control}

All heliostats at the NSTTF are operated using a single Master Control
System (MCS) based on an HP-1000 mini-computer.
Software to control heliostats and to track the sun
onto various targets has been previously developed by the NSTTF
staff.  The STACEE group has modified this software to track
astrophysical sources, accounting for the fact that air
showers originate near the top of the atmosphere and not at infinity.
The MCS is operated directly by STACEE scientists during observations

The MCS communicates with individual heliostats over a network of
serial lines.  Heliostat pointing coordinates are updated at least
once every 1.2 sec. 
Several video displays provide real-time
information on every aspect of the heliostat field and warn the
operator if there is a malfunction.  
Status and pointing information for every
heliostat is logged to a file every 5 seconds.

\section{Secondary Mirror Structure}

A schematic view of one of the three secondary mirror structures
is shown in Figure~3.
The structure consists of a commercial pallet stacker
which is secured to a steel base plate, or skid.
The stacker raises and lowers the mirror assembly 
consisting of a jib-crane and spider, both made of aluminum.
The jib-crane allows for 
azimuthal control by means of a linear 
actuator.
The spider holds the mirror and is attached to the jib-crane
by a ball joint and two actuators for elevation control.
The actuators are used only for alignment and the mirror
is not moved during observations.

\begin{figure}
\centerline{\psfig{file=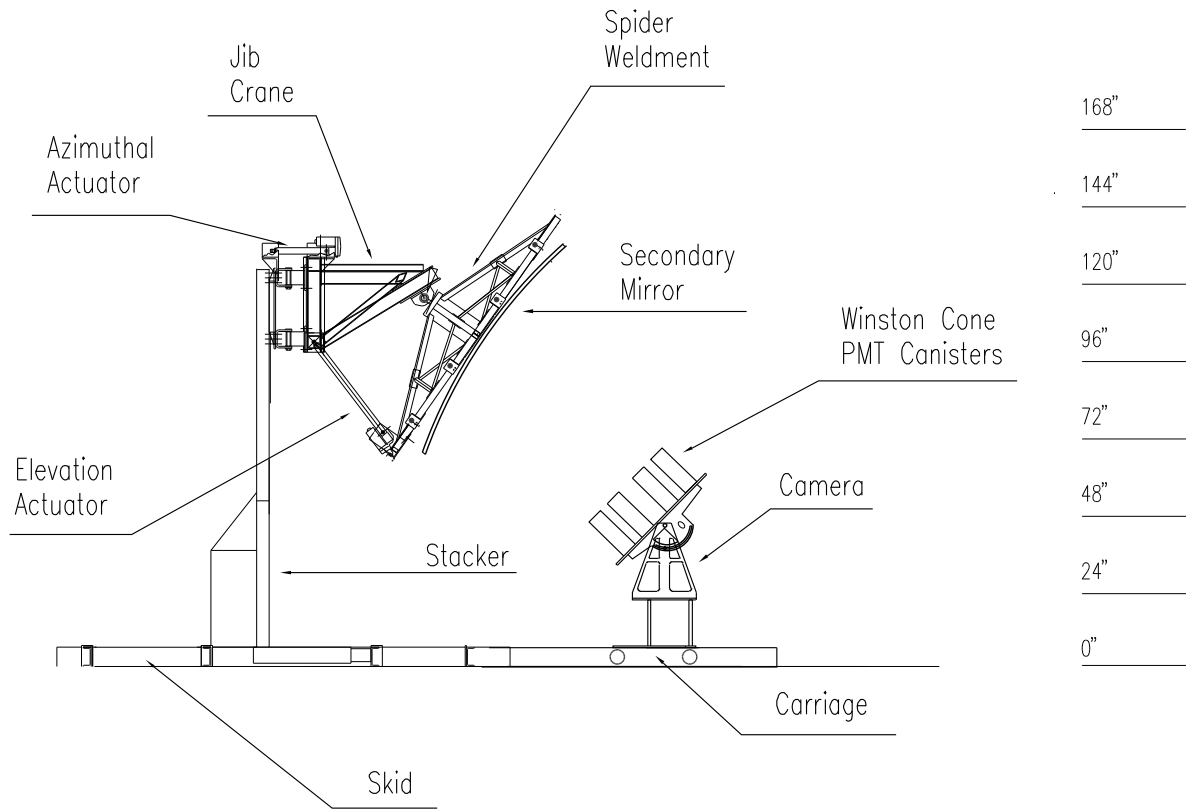,width=16.0cm}}
\caption{Schematic diagram of the STACEE secondary mirror
structure (one of three).
The structure supports a 2.0 m diameter spherical mirror with a
2.0 m focal length.
The mirror is positioned to reflect Cherenkov light from the
heliostat field into the PMT camera.
A full description of the various structure components is given
in the text.}
\end{figure}

\begin{figure}
\centerline{\psfig{file=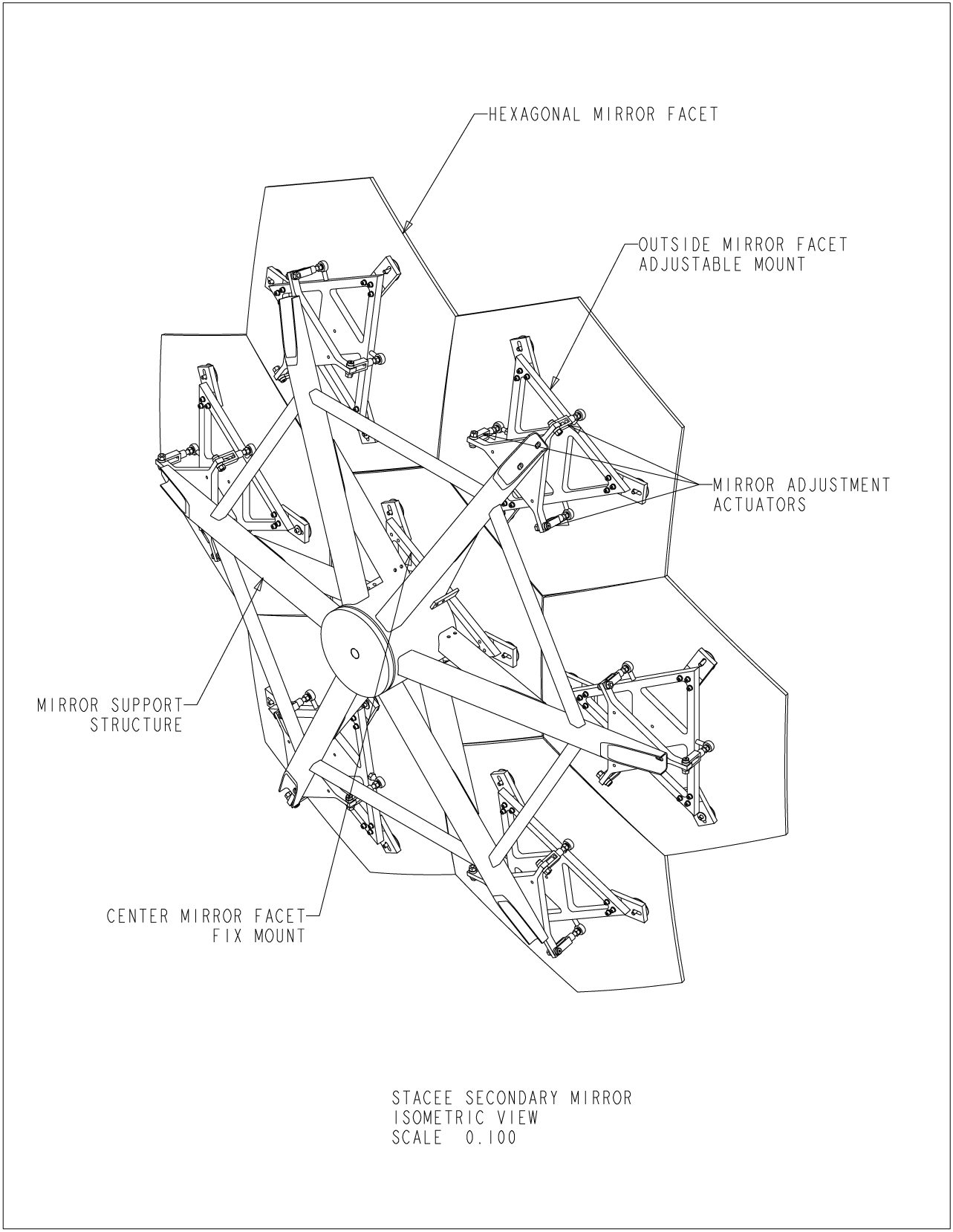,width=12cm}}
\caption{Schematic diagram of a STACEE secondary mirror
(one of three).
The mirror has an overall width of 2 m and a 2 m focal length.
It consists of seven hexagonal segments of ground, polished glass.
The segments are front-surfaced with aluminum and are mounted
to a rigid aluminum support structure known as the spider.
Alignment of each segment is accomplished by means of a
three point adjustment system, as shown.}
\end{figure}

The 2 m diameter mirror is spherical in shape and consists
of seven approximately hexagonal segments, as shown in Figure~4.
Each segment is a single piece of lime float glass
that has been slumped, ground, polished from a 0.5{\tt "} thick blank
to a 2 m focal length.
The glass is coated by evaporation of aluminum onto its front
surface and by a protective layer of AlSiO.
Each mirror segment is attached to the spider by means of
a triangular aluminum base-plate which is glued to the mirror
and is attached by bolts to the spider frame.
Mirror adjustments are made by a three-point mounting system on
each facet.
The overall mirror size is set by the
focal properties of the heliostats themselves.
Solar images have been measured for fifteen individual heliostats
and their sizes are reported in Covault {\em et al.} 1997.
The maximum image size at the tower is 1.7 m (FWHM)
and the image sizes scale as expected with the distance from the
tower.

%fuj
\newpage

\section{Camera}

\begin{figure}
\centerline{\psfig{file=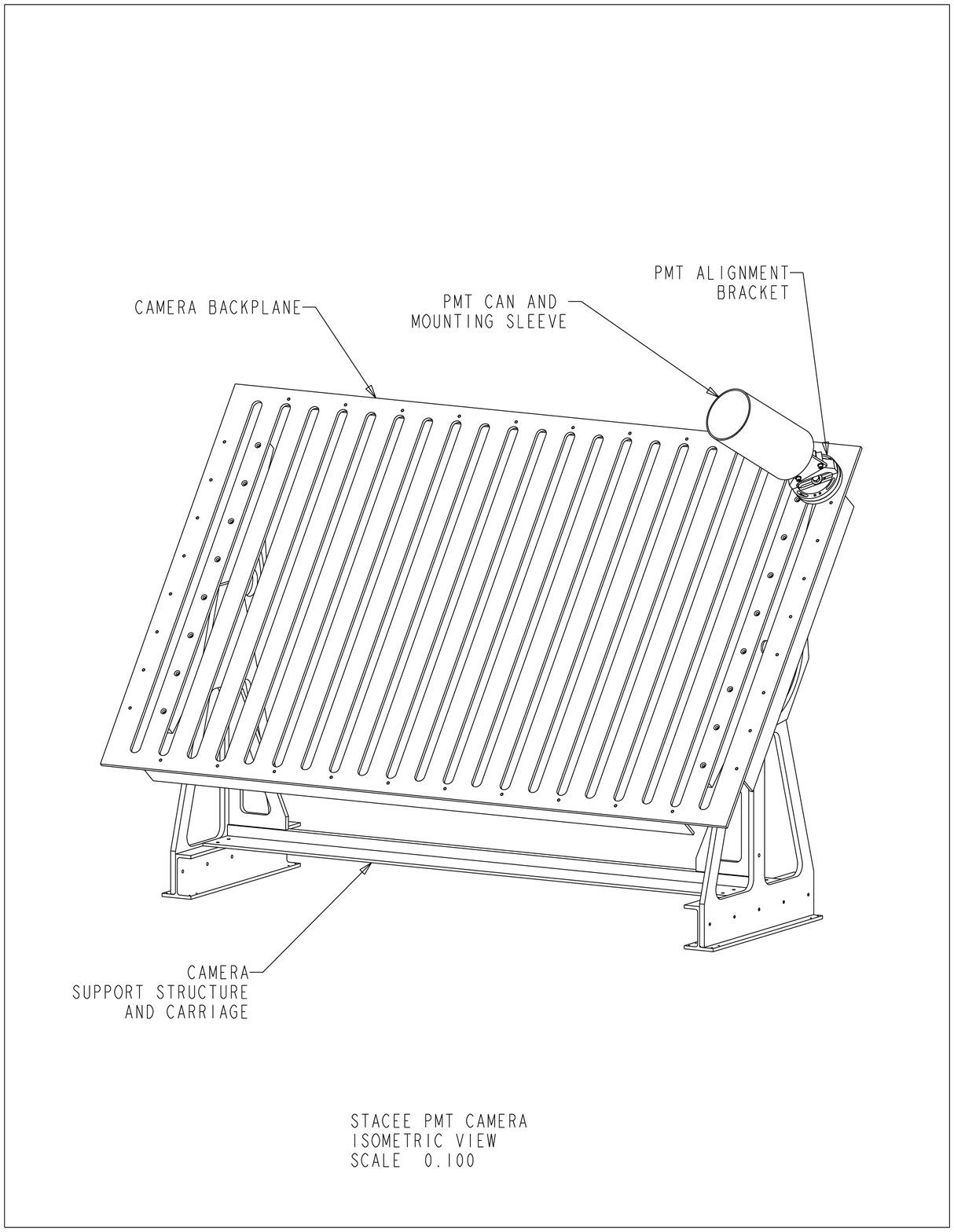,width=12cm}}
\caption{Schematic diagram of a STACEE camera (one of three).
The camera consists of a slotted aluminum backplane into
which PMT can assemblies are mounted.
Only one assembly (of sixteen) is shown for reasons of clarity.
A full description of the camera is given in the text.}
\end{figure}

The camera consists of two principal components:
an adjustable stand, and PMT can assemblies mounted on a
slotted plate, as shown in Figure~5.
The stand is a rigid structure made of aluminum 
bolted to a movable carriage.
The carriage allows the
entire camera to slide away from the tower edge to the proper
focal point of the secondary mirror.
The slotted plate 
is tilted in elevation to
allow for correct focusing of the
heliostat images in the center of the field-of-view of the mirror
onto the entrance apertures of their respective can elements.

Each PMT can element holds an assembly which comprises an acrylic
light concentrator followed by a photomultiplier tube.
The PMT and concentrator are held together in a hollow metal cylinder
which has an edge ring at
one end and a threaded disk at the other.
The disk pushes the concentrator and PMT together;
they are separated by a silicone rubber cookie which provides
good optical and mechanical coupling. 
This assembly is approximately 13 cm in diameter and 40 cm long.
The concentrators are dielectric total internal reflecting 
concentrators (DTIRCs) (Ning {\em et al.} 1987).

The can elements are mounted in a sleeve arrangement. 
The sleeves are secured to the slotted plate with a system that
allows for continuous x-y adjustment as well as canting so that 
PMTs away from the optical axis can be pointed towards the 
center of the mirror. 
For cans furthest from the optic axis of the mirror the
canting angle is 17$^\circ$. 
Operationally, one positions and cants the sleeves before 
sliding in the heavy PMT can elements. 

The DTIRC is approximated by a converging 
cylinder of 
%radius 11 cm at the
%front, radius 3.5cm at the back,
length between 10.6 cm and 13.2 cm.
The cylinder is
made of a single piece of UV transparent
acrylic that has been machined on a numerically controlled
mill to the proper dimensions.
The front surface of the DTIRC is a sphere 
and the rear surface is a circular aperture.
The DTIRC is designed so the field-of-view of each PMT is
restricted.
Three different types of DTIRCs are used corresponding to 
PMTs that view different parts of the heliostat field.
PMTs viewing heliostats closest to the tower are matched
with DTIRCs that have a field-of-view of 16$^\circ$
(half-angle), whereas PMTs
viewing heliostats farthest from the tower are matched
with DTIRCs having a field-of-view of 26$^\circ$.

The photomultiplier tubes are chosen for high speed and
moderate gain.
Speed is important to minimize the level of pile-up
from pulses created by night sky photons.
A moderate gain is needed so that a sufficient
amount of charge is generated, but at the same time
the total current draw of the tube must be kept to an
acceptable level ($< 100 \mu$A).
The PMTs (Philips XP2282) have
eight dynode stages and
a typical gain of $\sim 3\times 10^5$ at a high voltage
value of -1500 V.
Each tube has a transistorized base (Philips VD 182K/C)
that has been optimized for good linearity and speed.
The pulse width of a single photoelectron signal
at the output of the base is
$\sim 1.8$ nsec (FWHM) and the time transit spread of the pulse
is $\sim 0.45$ nsec.
The tube has a 2{\tt "} diameter bi-alkalai photocathode
and a front window made of borosilicate glass.

\section{Electronics}

The electronics serve two major purposes.
First, the PMT signals must be combined together in a coincidence
circuit to form the experiment trigger.
The trigger detects the occurrence of a Cherenkov air shower
amidst the fluctuations of the night sky background light.
Second, electronics are needed to measure the arrival time
and pulse-height of the Cherenkov signal reaching each of the PMTs.
The shower direction is reconstructed from the arrival time information
and the pulse-height information is used to estimate the shower energy.

\begin{figure}
\centerline{\psfig{file=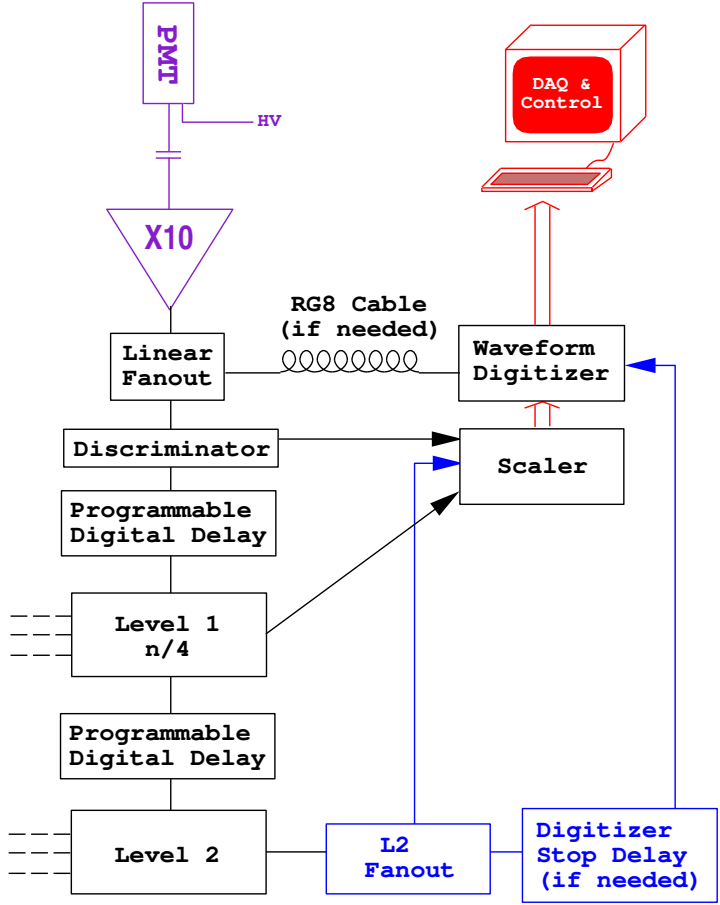,width=9.5cm}}
\caption{Block diagram of the STACEE electronics.
The PMT signals are amplified and discriminated.
The discriminated signals are delayed to form a
two-level (L1,L2) experiment trigger.
The trigger is used to gate fast digitizers which
record the PMT waveforms.}
\end{figure}

A block diagram of the electronics is shown in Figure~6.
Commercial components are used wherever possible.
The PMTs are powered by a high voltage mainframe (Lecroy 4032).
The PMT signals are capacitively coupled (time constant = 75 nsec),
amplified (Phillips 776), and discriminated (Lecroy 4413).
Copies of the PMT signals are generated by a linear fan-out
(Phillips 748) and these copies are sampled by waveform digitizers.
The waveform information is read out via Ethernet by the
data acquisition (DAQ) computer.

The trigger is formed by a coincidence of the discriminated PMT signals
that are delayed appropriately to bring the signals into time.
The delays account for the fact that the Cherenkov light reaches
different PMTs at different times because of variations in the times of
flight from individual heliostats to
the secondary mirror.
In addition, delays are needed to account for time of flight differences
due to the orientation of the shower wavefront.
The trigger is composed of two distinct parts: Level 1 (L1) and
Level 2 (L2)

The discriminated signals are delayed
by programmable delay units (Lecroy 4518) and fixed length time of
flight cables (RG 58U) and combined to form
the L1 trigger.
The L1 trigger consists of multiplicity logic
acting on parallel groups of four PMTs (i.e. N of 4 logic).
The PMTs are grouped to form L1 triggers based on
the proximity of their respective heliostats.
The L1 trigger signals are passed through additional programmable
delay (CAEN 469) to the L2 multiplicity logic, which makes
a majority decision based on the number of in-time L1 signals
(i.e. N of 12 logic).
The L2 trigger signifies an event trigger which
stops the waveform digitizers and initiates an event read-out.

The waveform digitizers are 1 GHz sampling units with 8-bit
resolution.  
Two commercially available products are being considered.
One is a digitizer made in the VXI standard (Tektronix TVS641)
and the other is a flash ADC in the VME standard (ETEP 301).

The DAQ computer will be a Silicon Graphics Workstation
(Indy R4400 or equivalent).
The workstation will control the CAMAC components via an
Ethernet crate controller
(Hytec 1365).
Universal time will be recorded via a
CAMAC-based Global Positioning System (GPS) clock
(Hytec GPS92).

%\section{Other Components}

\section{Conclusion}

STACEE is an experiment designed to carry out gamma-ray observations
in the unopened window between 20 and 250 GeV using the
atmospheric Cherenkov technique.
The experiment uses large heliostat mirrors of the National
Solar Thermal Test Facility 
to achieve a very large Cherenkov photon collection area.
Additional components of STACEE include secondary mirrors and
structures, cameras, and electronics.
The baseline design for STACEE has been finalized and
the full experiment is now under construction.

\vspace{10mm}

{\bf \noindent Acknowledgments}

\vspace{5mm}

STACEE is a collaboration of scientists from
the University of Chicago, McGill University,
the University of California (Santa Cruz),
the University of California (Riverside),
Barnard College, Yale University, and
California State University (Los Angeles).
We acknowledge the help of E. Pod, M. Houde,
the engineering staff of the
Yerkes Observatory, and the personnel of the McGill Physics Department
machine shop.
We wish to thank P. Fleury, E. Par{\' e}, J. Qu{\' e}bert, and D.A. Smith
for many useful discussions.
This work is supported in
part by the National Science Foundation, the Institute of Particle
Physics of Canada, the Natural Sciences and Engineering Research
Council, and the California Space Institute.  RAO acknowledges the
support of the Grainger Foundation and the 
University of Chicago.
CEC is a Cottrell Scholar of the Research Corporation.

\begin{refs}

\item Chantell, M. {\em et al.,}
Nucl Inst. Meth. Phys. Res., in press (1997).

\item Covault, C. {\em et al.,}
these proceedings (1997).

\item Danaher, S. {\em et al.,}
Solar Energy 28, 355 (1982).

\item Ning, X., Winston, R., and
O'Gallagher, J.,
Applied Optics 26, 300 (1987).

\item NOAA, National Climatic Data Center,
U.S. Department of Commerce,
Local Climatological Data, Albuquerque NM (1997).

\item Ong, R.A. {\em et al.,}
Astroparticle Phys. 5, 353 (1996).

\item Ong, Rene A.,
{\em Very High Energy Gamma-Ray Astronomy},
EFI 97-41, Physics Reports, submitted (1997).

\item Par{\' e}, E.
these proceedings (1997).

\item Plaga, R.,
these proceedings (1997).

\item O.T. T{\" u}mer {\em et al.,}
Nucl. Phys. B (Proc. Suppl.) 14A, 351 (1990).
\end{refs}

\end{document}